\documentclass[
aps,
prl,
reprint,
floatfix,
amsmath,
twocolumn,
superscriptaddress,
altaffillsymbol
]{revtex4-1}

\usepackage{dcolumn}   % needed for some tables
\usepackage{bm}        % for math
\usepackage{amssymb}   % for math
\usepackage{amsfonts}
\usepackage{verbatim}
\usepackage{mathrsfs}
\usepackage{graphicx}
\usepackage{amsmath}
\usepackage{amsfonts}
\usepackage{ulem}

\newcommand{\ket}[1]{|#1\rangle}

\def\lsim{\mathrel{\rlap{\lower4pt\hbox{\hskip1pt$\sim$}}
    \raise1pt\hbox{$<$}}}                % less than or approx. symbol
\def\gsim{\mathrel{\rlap{\lower4pt\hbox{\hskip1pt$\sim$}}
    \raise1pt\hbox{$>$}}}                % greater than or approx. symbol

\begin{document}

\title{Quantum--Classical Interface Based on Single Flux Quantum Digital Logic} %Title of paper

\author{R. McDermott}
%\thanks{These authors contributed equally to this work. T. T. is currently at IBM T. J. Watson Research Center, Yorktown Heights, New York 10598, USA.}
\email[Electronic address: ]{rfmcdermott@wisc.edu}
\author{M. G. Vavilov}
%\thanks{These authors contributed equally to this work. T. T. is currently at IBM T. J. Watson Research Center, Yorktown Heights, New York 10598, USA.}
\affiliation{Department of Physics, University of Wisconsin-Madison, Madison, Wisconsin 53706, USA}
\author{B. L. T. Plourde}
\affiliation{Department of Physics, Syracuse University, Syracuse, New York 13244, USA}
\author{F. K. Wilhelm}
\author{P. J. Liebermann}
\affiliation{Theoretical Physics, Saarland University, 66123 Saarbr\"{u}cken, Germany}
\author{O. A. Mukhanov}
\affiliation{HYPRES, Inc., Elmsford, NY 10523, USA}
\author{T. A. Ohki}
\affiliation{Raytheon BBN Technologies, Cambridge, Massachusetts 02138, USA}

%\author{T. Thorbeck}
%\thanks{These two authors contributed equally.}
%\altaffiliation[Present address: ]{IBM T. J. Watson Research Center, Yorktown Heights, New York 10598, USA}
%\affiliation{Department of Physics, University of Wisconsin-Madison, Madison, Wisconsin 53706, USA}
%%\affiliation{Department of Physics, University of Wisconsin-Madison, Madison, Wisconsin 53706, USA}
%\author{S. Zhu}
%\thanks{These two authors contributed equally.}
%\affiliation{Department of Physics, University of Wisconsin-Madison, Madison, Wisconsin 53706, USA}
%\author{E. Leonard Jr.}
%\affiliation{Department of Physics, University of Wisconsin-Madison, Madison, Wisconsin 53706, USA}
%\author{R. McDermott}
%\email[Electronic address: ]{rfmcdermott@wisc.edu}
%\affiliation{Department of Physics, University of Wisconsin-Madison, Madison, Wisconsin 53706, USA}

%\email[Electronic address: ]{rfmcdermott@wisc.edu}
%\affiliation{Department of Physics, University of Wisconsin-Madison, Madison, Wisconsin 53706, USA}
%\altaffiliation[Present address: ]{IBM T. J. Watson Research Center, Yorktown Heights, New York 10598, USA}

\date{\today}

\begin{abstract}

We describe an approach to the integrated control and measurement of a large-scale superconducting multiqubit circuit using a proximal coprocessor based on the Single Flux Quantum (SFQ) digital logic family. Coherent control is realized by irradiating the qubits directly with classical bitstreams derived from optimal control theory. Qubit measurement is performed by a Josephson photon counter, which provides access to the classical result of projective quantum measurement at the millikelvin stage. We analyze the power budget and physical footprint of the SFQ coprocessor and discuss challenges and opportunities associated with this approach.

\end{abstract}

\pacs{}% insert suggested PACS numbers in braces on next line

\maketitle %\maketitle must follow title, authors, abstract and \pacs

\section{I. Introduction}

Superconducting quantum circuits are a leading candidate for scalable quantum information processing \cite{Schoelkopf08, Clarke08, Gambetta17}. Gate and measurement fidelities are at the threshold for fault tolerance in the two-dimensional surface code \cite{Kelly15} and there is interest in scaling to larger systems. However, the hardware overhead associated with the surface code is immense: a practical factoring machine is expected to require about 100 million physical qubits \cite{Fowler12}, far beyond current capabilities. While brute-force scaling with current technology might be adequate to realize qubit arrays of order 100 qubits \cite{Boixo16}, it is unknown how to scale superconducting quantum circuits to the thousands, much less millions, of physical qubits required to realize a large-scale quantum array. The surface code requires high-fidelity entangling operations between nearest neighbors, in addition to high-fidelity single qubit gates across the physical qubit array and high-fidelity measurement on at least half the array. Recent progress in three-dimensional integration points a direction to the realization of large-scale qubit arrays with the required connectivity \cite{Rosenberg17, Foxen17}, and it is likely that such arrays can be engineered in a manner to preserve error rates at levels well below threshold. For current technology based on pulsed microwave control and amplification followed by heterodyne detection, however, the heat load and physical footprint of the required classical hardware preclude scaling to qubit arrays approaching $10^6$ elements. The implementation of a scalable classical coprocessor for control and error tracking of the quantum array represents one of the key challenges facing the community.  This challenge goes far beyond the realm of ``mere" engineering, as continued progress will require the development of new technologies and approaches for both coherent control and measurement. 

Currently, control in superconducting qubits is accomplished via shaped microwave tones that realize arbitrary rotations over the Bloch sphere. Amplitude modulation of a resonant carrier wave concentrates drive power at the frequency of interest, and pulses are shaped to minimize power at nearby transition frequencies to avoid excitation out of the qubit manifold \cite{Motzoi09, Motzoi11}. The microwave control requires one low-phase noise generator, a quadrature mixer, and two high-speed DAC channels at room temperature to generate rotations with arbitrary amplitude and phase on the Bloch sphere. Two-qubit gates are accomplished via coupling through a linear bus \cite{Majer07} or microwave cross resonance \cite{Rigetti10, Chow11, Chow13} or via shaped flux pulses that exploit an avoided level crossing between the $\ket{11}$ and $\ket{20}$ states \cite{Strauch03, DiCarlo09}, requiring a separate high-speed DAC channel. In an arrangement such as the surface code where the logical qubits are formed from a periodic array of physical qubits, it is possible to ``recycle" frequencies, keeping the number of required microwave tones to a minimum \cite{Helmer09, Versluis17}. What is less clear is whether it is possible to recycle pulse waveforms while maintaining high gate fidelity, as errors depend sensitively on pulse shape, and analog waveforms are susceptible to distortion and losses that in general will vary from one control channel to the next. In the traditional control paradigm using shaped analog pulse waveforms, bringup of each singe- and two-qubit gate is a separate optimization problem.

Qubit measurement is conventionally accomplished via microwave heterodyne detection. In the circuit quantum electrodynamics (circuit QED) architecture \cite{Blais04, Wallraff04}, the qubit is dispersively coupled to a linear resonator and interaction between the two modes imparts a qubit state-dependent frequency shift on the resonator. It is therefore possible to probe the qubit state by monitoring microwave transmission across or reflection from the resonator. In a typical measurement configuration, the microwaves scattered from the qubit readout resonator are first amplified by a near quantum-limited amplifier followed by postamplification by a High Electron Mobility Transistor (HEMT) and subsequent heterodyne detection and thresholding at room temperature. In the context of the surface code, error detection demands fast, high-fidelity measurement of multiqubit parity operators \cite{Fowler12}. In the most usual approach, the parity bit is read out using an ancilla qubit, although various approaches to direct parity measurement have been pursued \cite{DiVincenzo13, Riste13}. To achieve high fidelity, it is necessary that the noise contribution of the first-stage amplifier be close to the standard quantum limit. A variety of ultralow-noise Josephson amplifiers have been applied to the high-fidelity measurement of superconducting qubits \cite{Vijay11}; however, the demands of operating a large-scale superconducting processor require global optimization of the measurement chain, and amplifier added noise is but one consideration. Multiplexed readout requires both large instantaneous bandwidth and high saturation power of the first stage amplifier, and several amplifiers show promise as the first gain stage in a multiplexed qubit measurement system, including the Impedance-Matched Parametric Amplifier (IMPA) \cite{Mutus14}, the Traveling-Wave Parametric Amplifier (TWPA) \cite{Macklin15}, the Kinetic Inductance Traveling-wave amplifier (KIT) \cite{Eom12, Vissers16}, and the Superconducting Low-inductance Undulatory Galvanometer (SLUG) \cite{Hover14}.  Furthermore, the measurement system must isolate the qubit from the noise of downstream amplification stages at higher temperatures while at the same time producing minimal classical backaction on the qubit, due either to stray microwave power from pump tones or to emission from dissipative elements. For this reason, it is generally necessary to incorporate nonreciprocal elements between the qubit and downstream measurement stages. Commercial ferrite-based isolators and circulators are bulky, magnetic, and expensive, so they are not a scalable technology. There have been prior attempts to engineer nonreciprocal gain in superconducting parametric amplifiers, notably using coupled Josephson parametric converters (JPCs) \cite{Abdo13, Abdo14}. However, the bandwidth and saturation power of such devices are quite limited, complicating efforts to perform multiplexed qubit readout. Given the current state of technology, the hardware footprint of the amplifiers, cryogenic isolators, and room-temperature electronics required for heterodyne detection and thresholding is immense, and the path to scalability is unclear.

For a scalable system, it is highly desirable to integrate as much of the control and measurement circuitry as possible in the multiqubit cryostat in order to reduce wiring heat load, power consumption, and the overall system footprint, and to allow for low-latency feedback for error correction. An obvious candidate for the cold control system is Single Flux Quantum (SFQ) digital logic, in which classical bits of information are stored in propagating fluxons, voltage pulses whose time integral equals the superconducting flux quantum $\Phi_0 = h/2e$ \cite{Likharev91, Bunyk01a}. For classical digital and mixed-signal applications, SFQ circuits have achieved relative maturity; notable accomplishments include the realization of complex digital processing circuits \cite{Filippov12, Tang16, Sato17} and practical wideband receiver systems \cite{Mukhanov08}. However, the development of SFQ-based classical logic circuits for qubit control and measurement has proceeded slowly (see Section II below). Our team has recently proposed a new scheme for coherent quantum control using resonant SFQ pulse trains \cite{McDermott14}. We have analyzed the fidelity of SFQ-based gates both analytically and using Monte Carlo simulations, and we have shown that these gates are robust against leakage errors and timing jitter of the pulses, with achievable fidelities in excess of 99.9\% in gate times around 20~ns. Investigations by some of us demonstrate that superconducting quantum circuits can be made robust against the inevitable quasiparticle poisoning that will come with an integrated SFQ pulse driver \cite{Patel16}, and preliminary experiments have been performed to demonstrate coherent qubit control with resonant SFQ pulse trains \cite{Leonard17}. While leakage out of the computational basis will ultimately limit the fidelity of naive, resonant SFQ-based control sequences, it has been shown that by appropriate variation of the pulse-to-pulse interval in the control sequence, gate errors can be suppressed by 2 orders of magnitude or more \cite{Liebermann16}.

Just as it is possible to coherently control a qubit array using quantized digital logic pulses, it is possible to map the outcome of quantum measurement to a classical bit that is accessible at the millikelvin stage of the cryostat, so that it can be exploited for low-latency quantum feedback and control conditioned on the result of qubit measurement. Our team has proposed an efficient qubit measurement scheme that involves encoding the qubit state to microwave cavity pointer states \cite{Govia14, Govia15}. In this case, qubit measurement can be achieved by coupling the readout resonator to a Josephson microwave photon counter \cite{Chen11, Walsh17}. We have performed a preliminary version of the microwave counter-based measurement protocol and demonstrated raw single-shot measurement fidelity around 92\% \cite{Opremcak17}. We believe with straightforward refinements of the measurement protocol that it will be possible to achieve single-shot measurement fidelity around 99\%. Crucially, the classical binary output of the counter can be easily converted to a propagating fluxon suitable for postprocessing by a proximal SFQ-based classical controller. The notion of SFQ-based coherent control, taken together with a scheme for high-fidelity qubit measurement with a photon counter, points a direction toward the integration of large-scale superconducting quantum circuits with proximal control and measurement circuitry based on SFQ digital logic.
\begin{figure}[t]
\includegraphics[width=.98\columnwidth]{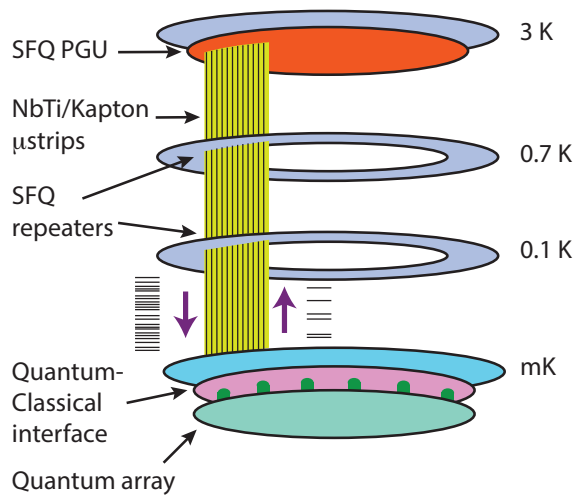}
\vspace*{-0.0in} \caption{Scheme for SFQ-based classical coprocessor for control and error tracking of a large scale quantum array. The Pattern Generator Unit (PGU) at the 3~K stage of the cryostat stores and streams dense classical bitstreams to the quantum array to induce coherent rotations and entangling gates. The dilute results of projective quantum measurement are streamed upward from the quantum array to the SFQ coprocessor. The interface layer at the millikelvin stage mediates the interaction between the quantum array and the classical coprocessor. Communication between the classical coprocessor and the interface layer is accomplished via superconducting microstrip flex lines, with SFQ repeater stages at intermediate temperatures to ensure accurate timing and faithful transmission of classical bitstreams.}
\label{fig:vision}\end{figure}

Our vision of an integrated SFQ-based classical coprocessor for the control and monitoring of a large-scale superconducting quantum computer is shown in Fig. \ref{fig:vision}. The quantum circuit resides at the millikelvin stage of a large-capacity dilution refrigerator; we assume an available cooling power at this stage of order 10~mW. For the sake of concreteness, we consider a two-dimensional array of $10^8$ qubits with nearest-neighbor coupling and local control and measurement. The footprint per physical qubit cell is $100\times100$~$\mu$m$^2$, corresponding to a footprint of 1~m$^2$ for the array as a whole. The classical coprocessor incorporates both SFQ-based pattern generator units (PGUs) that are used to create digital bit patterns for qubit control, as well as logic units used to process the results of counter-based qubit measurement for the purpose of error tracking and, if necessary, to provide low-latency feedback to stabilize the quantum array. The classical coprocessor resides at the 3~K stage of the qubit cryostat, where we assume an available cooling power of order 10~W. Digital pulse patterns will be streamed to the quantum array over low-loss superconducting microstrip flex lines, and the dilute results of stabilizer measurements will be streamed upward to the coprocessor. If necessary, SFQ repeater stages will be located at intermediate temperatures to ensure the high-fidelity communication of classical information between the quantum array and the coprocessor. Crucial to the success of the scheme is the existence of an interface layer at the millikelvin stage to provide for high-fidelity communication of bit patterns and measurement results across the quantum--classical divide. The interface chip will incorporate SFQ pulse drivers; photon counters with integrated SFQ converters for the transmission of measurement results upwards to the coprocessor; and SFQ-based multiplex/demultiplex (MUX/DEMUX) elements to minimize the wire count needed for communication of classical bit streams between the interface chip and the coprocessor.  The interface chip will be coupled to the quantum array in a flip-chip arrangement; coupling between the interface chip and the quantum array will be accomplished capacitively and inductively, with no need for galvanic transmission of signals between the two chips.

Our scheme offers a number of advantages for robust coherent control of large-scale quantum circuits:

\begin{itemize}

\item{First, the implementation of proximal cryogenic control hardware is a prerequisite to realization of a scalable system. It is critical to maintain a slender profile in terms of both hardware footprint and power consumption throughout the measurement and control stack. The ability to integrate much of the classical processing at low temperature allows a dramatic reduction in wire count and heat load from 300~K to 3~K and greatly reduces the hardware demands at room temperature. The proposed implementation is well-matched to the cooling power and experimental space available from a large-capacity, special-purpose dilution refrigerator cryostat.}

\item{Second, integration of the coprocessor at the millikelvin stage offers the possibility of low-latency feedback for stabilization of the quantum array, or for monitoring and correction of leakage errors. Prior attempts to stabilize arbitrary quantum states have been constrained by the significant time delay associated with signal amplification, heterodyne detection, thresholding, and conditional control with room temperature electronics \cite{Vijay12, Riste15}. The ultrafast clock speed of the SFQ coprocessor and the proximity of the classical decision engine to the quantum array offer distinct advantages.}

\item{Third, our approach will enable smart system identification for calibration and bringup of quantum circuits. The response of the qubit to well-defined SFQ bitstreams will provide a fingerprint of the device that will allow us to extract the qubit 01 transition frequency and higher transitions in an efficient manner.}

\item{Finally, we are proposing to move quantum control from the analog realm to the digital realm, and all of the robustness associated with digital control in the classical regime will carry over to the quantum regime. In our implementation, the size of a qubit rotation is determined entirely by one geometric coupling parameter and by the size of the magnetic flux quantum, a fundamental constant of Nature. To a very good approximation, the geometric coupling of control lines to the qubit does not fluctuate, and every magnetic fluxon is ``perfect'' and identical. Our SFQ-based pulse sequences are thus immune to distortion due to unknown parasitics in the control wiring. Once system identification is accomplished, the control problem can be fully understood and robust solutions can be tailored that are immune to phase noise of cabling, long-term gain drifts in DAC controllers, etc.}

\end{itemize}

Below we propose one approach to realization of an SFQ-based coprocessor for qubit control and measurement. This manuscript is organized as follows. In Section II we provide a historical perspective on prior attempts to marry SFQ classical logic with qubit circuits. In Section III we describe recent developments in the area of ultralow power SFQ logic and discuss the power consumption of SFQ elements operated at various stages of the multiqubit cryostat. In Section IV we discuss in detail our proposed approach to SFQ-based qubit control with a focus on achievable gate fidelity for realistic device parameters. Section V provides an introduction to photon counter-based qubit measurement. We describe a possible implementation of the SFQ-based PGU for qubit control in Section VI. Section VII includes estimates of the power consumption and physical footprint of the SFQ coprocessor and interface array, along with a discussion of the requirements for wiring heat load and connectivity between the subsystems. Finally, in Section VIII we conclude and discuss challenges and opportunities associated with realization of a scalable quantum--classical interface.

\section{II. Historical Overview}

\indent \indent The first proposals for monolithic integration of qubits and SFQ circuits were focused on the demonstration of macroscopic quantum coherence \cite{Feldman00, Rey01}; however, these works also explored possible approaches to SFQ-based qubit control and measurement \cite{Zhou01, Crankshaw03, Semenov03}. Early experiments involved complex circuits with large critical currents and on-chip bias resistors, which were a source of excess power dissipation and heating at millikelvin temperatures. Moreover, little to no effort was made to preserve high quantum coherence with the introduction of a dissipative quantum--classical interface. Subsequent work focused on the thermal budget \cite{Ohki05, Ohki05b, Intiso06, Savin06} and electromagnetic compatibility \cite{Rott03, Zorin05, Hassel06, Ohki07} of the SFQ elements, critical considerations for minimizing decoherence. During the European project RSFQubit, a foundry was established at VTT \cite{Gronberg07} that could provide some unique features required for millikelvin operation of SFQ elements, including critical current densities from 10-30~A/cm$^2$, Cu cooling fins for thermalization of shunt resistors \cite{Wellstood94}, and quasiparticle traps.  In the US, HYPRES, Inc. also offered a low-current density process for monolithic SFQ integration with qubits. However, Nb-based qubits fabricated with these SFQ processes displayed poor coherence. This is partly due to the low intrinisic quality factor of the  SiO$_2$ wiring dielectric \cite{Martinis05}, but also to the Nb-AlO$_x$-Nb trilayer junction process, which has never produced high-quality superconducting qubits. While in many cases the amplitude and timing resolution of the SFQ controller were insufficient to allow for high-fidelity qubit control, over time sophisticated SFQ-based approaches to baseband control of large qubit arrays were developed, notably by DWave \cite{Johnson10}. With respect to readout, several approaches were developed for the detection of the flux state of a superconducting loop \cite{Ohki07b, Herr07}; these could be applied in a straightforward way to the measurement of flux or phase qubits.

Two developments starting around the year 2005 significantly altered the direction of the superconducting qubit field, ultimately forcing a retreat from the early ambitious efforts at SFQ--qubit integration. First, it was realized that two-level state (TLS) defects in the materials used to realize the qubit constituted a major source of decoherence \cite{Martinis05}, prompting a focus on simple, stripped-down fabrication processes based on double-angle evaporation of Al-AlO$_x$-Al Josephson junctions. Around the same time, circuit QED \cite{Blais04, Wallraff04} emerged as an extremely powerful paradigm for the operation and measurement of superconducting qubits. The complex multilayer fabrication processes developed earlier with an eye to SFQ--qubit integration were not suited to the realization of high-coherence qubits, and the early ideas for SFQ-based flux detection did not target the needs of the dispersive microwave readout schemes used in circuit QED. In the end, the superconducting qubit field progressed rapidly and the idea of monolithic integration of an SFQ coprocessor with the qubit circuit was left behind.

In the view of these authors, the notion of an SFQ-based coprocessor to support a large-scale superconducting quantum computer was not fundamentally flawed, but rather out of sync with the qubit technology of the time: the development of highly coherent qubit arrays of course needed to precede any serious effort to develop scalable approaches to control and measurement. Today we have a much firmer understanding of the limits to qubit coherence. Qubit gate and measurement fidelity have attained the fault-tolerant threshold \cite{Barends14, Corcoles15}, and in order to realize large-scale qubit arrays it is necessary to move beyond the simple, stripped-down circuits suitable for initial demonstrations to complex hybrid circuits involving multichip modules (MCMs) \cite{Rosenberg17, Foxen17}. At the same time, there has been significant progress toward the development of ultralow-power variants of SFQ digital logic, opening the possibility of tight integration of SFQ elements with superconducting quantum circuits at the millikelvin stage; we describe these developments in detail below.

\section{III. Ultralow Power SFQ Logic}

\indent \indent The power consumption of the SFQ coprocessor must be kept at a minimum for integration with a multiqubit circuit; however, conventional Rapid Single Flux Quantum (RSFQ) logic \cite{Likharev91} relies on a resistor-based dc current bias network that is responsible for the dominant static part of the total power dissipation. In fact, Joule heating in the bias network exceeds the fundamental dynamic dissipation associated with SFQ processing by a factor of 60-70 \cite{Mukhanov11}. As a result, conventional RSFQ logic is ill-suited to the realization of very large scale integration (VLSI) circuits or implementation of a scalable quantum--classical interface. Fortunately, new energy efficient post-RSFQ technologies have been introduced in which the dominant static contribution to power dissipation has been eliminated \cite{Mukhanov11, Kirichenko11, Herr11, Volkmann13, Takeuchi13}. Broadly speaking, these new approaches exploit dissipationless inductive dividers to distribute bias to the various parts of the SFQ processor \cite{Mukhanov11}, or they involve ac bias schemes that move the dissipation of the bias source off chip \cite{Herr11}. Among these new logics, ERSFQ and eSFQ are the closest to conventional RSFQ, with the majority of logic gates shared between RSFQ and these low-power successors \cite{Mukhanov11, Kirichenko11, Volkmann13}. For the low-power variants of SFQ logic, power dissipation is determined only by the energy per phase slip and the circuit clock speed $f_{\rm clk}$: $P = \Phi_0 I_b f_{\rm clk}$, where the dc bias current $I_b$ is typically $\sim$75\% of the gate critical current $I_c$.  The typical critical current for an SFQ junction designed for 3-4~K operation is 100~$\mu$A, set by the requirement that gates remain robust against thermal fluctuations. This critical current corresponds to an SFQ switching energy per junction of order $10^{-19}$~J and a power dissipation of order 1~nW for an average phase slip rate of 5~GHz.  However, it has been observed that even 1~nW of power dissipated locally on chip can increase the electrical temperature by 10s of mK from a substrate temperature of 30~mK \cite{Intiso06}.  As a result, both ERSFQ-type bias distribution schemes and drastic reductions of $I_c$ must be used to enable monolithic integration of a quantum array with a large-scale SFQ circuit.

\begin{figure}[t]
\includegraphics[width=.98\columnwidth]{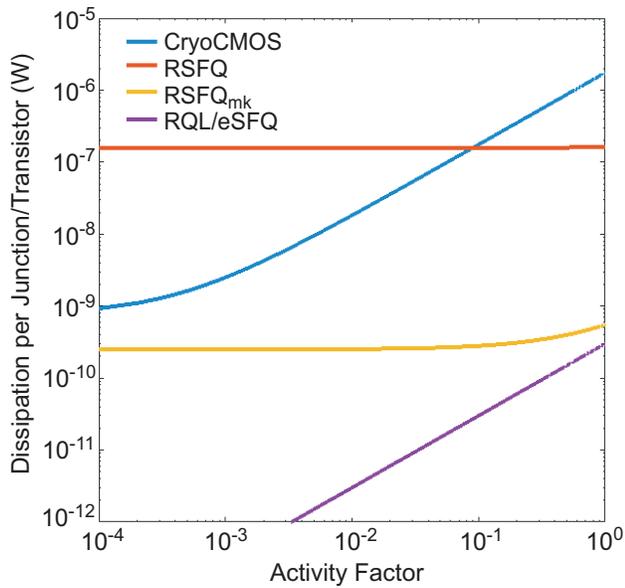}
\vspace*{-0.0in} \caption{Comparison between CMOS and SFQ for various implementations and parameters.  CryoCMOS is a potential 4~K CMOS technology with reduced $V_{\rm dd}$; RSFQ is conventional 4~K SFQ digital logic technology; RSFQ$_{\rm mK}$ refers to low-$I_c$, millikelvin-optimized RSFQ; and RQL/ERSFQ are low-power SFQ variants that eliminate bias resistors and therefore entail negligible static power dissipation.}
\label{fig:TAO}\end{figure}

Typical $I_c$ values for millikelvin-compatible SFQ circuits can be 100 times smaller than those for circuits designed for operation at 4~K; SFQ junctions with $I_c$ of order 1~$\mu$A will still remain robust against thermal fluctuations at dilution refrigerator temperatures. The  associated switching energy is of order $10^{-21}$~J, orders of magnitude lower than that of cryogenic CMOS. In Fig. \ref{fig:TAO} we plot the power dissipation for several variants of CMOS and SFQ digital logic \textit{versus} activity factor; here we assume a 10~GHz clock frequency and activity factor is defined as the fraction of clock cycles during which the logic element switches (i.e., an activity factor of 1 corresponds to a switching event in each clock cycle).  The power was calculated by assuming an even mix of logic to memory gates and a maximum activity factor of 1 for logic and 0.5 for memory.  For CMOS, the $V_{dd}$  used was 0.5~V and the effective capacitance for the gates was 0.5~fF/$\mu$m.  A specific logic and memory device capacitance was derived from typical devices and an estimated 4~K gate leakage of 1.5~nA was used to calculate the static power consumption.   For the SFQ circuits, the $I_c$ used for calculation was 250, 10, and 10~$\mu$A for RSFQ, RSFQ$_{\rm mK}$ and RQL/ERSFQ, respectively.  For RSFQ and RSFQ$_{\rm mK}$, the bias resistors responsible for static dissipation were on the order of the typical shunt resistors for the Josephson junctions in the circuits.  Generally in translating a classical logic gate from CMOS to SFQ, the resulting Josephson junction count is not equal to the transistor count of the original circuit.  Depending on the complexity of the gates, this ratio must be taken into account when comparing the normalized dissipation per device for SFQ and CMOS.

One fundamental constraint that leads to a potential scaling problem is that the $L I_c$ product of the SFQ cell must be maintained at roughly $0.5 \, \Phi_0$ for transmission lines and $\sim 1 \, \Phi_0$ for many gates.   A side effect of the reduction in $I_c$ is a corresponding increase of the cell inductance, which directly scales the cell area.  Moreover, the inductor does not act as a lumped element once its length approaches the wavelength for the propagating pulses. For SFQ junctions with $I_c \lsim 8 \, \mu$A, the length of the storage loop inductor exceeds this limit for conventional thin-film inductor technology. One solution to this problem is to utilize a geometrically short inductor provided by supplemental series Josephson junctions or by a high kinetic inductivity nanowire \cite{Goltsman01}.

Despite recent advances, there is a big leap to be taken for SFQ technologies to achieve complexity and integration density on par with mature CMOS.  The fundamental tension between power dissipation and physical footprint exacerbates the problem of low integration density for SFQ circuits tailored for millikelvin operation. It is critical to develop streamlined circuit solutions that minimize the number of SFQ gates required for qubit readout, error correction, and control functions, as opposed to recapitulating CMOS circuits using SFQ technology.

\section{IV. SFQ-based Coherent Control}

\begin{figure}[t]
\includegraphics[width=.9\columnwidth]{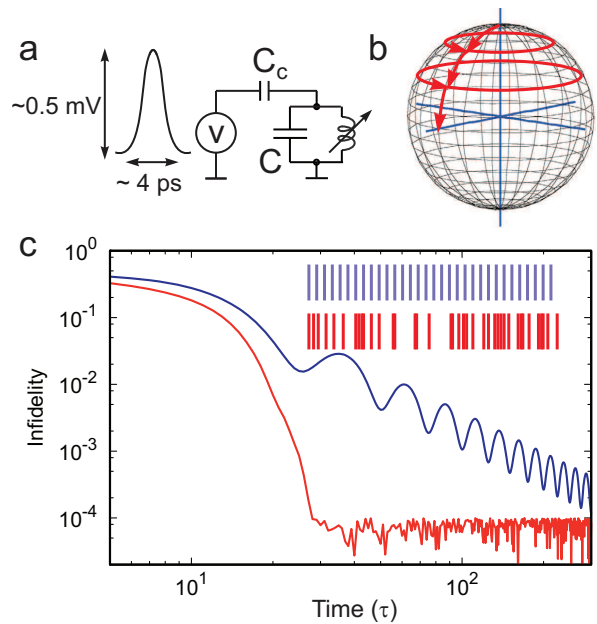}
\vspace*{-0.0in} \caption{(a) Excitation of a resonant mode via a train of SFQ pulses. The pulses are coupled to the resonator through the capacitance $C_c$. For 4.5 kA/cm$^2$ Nb-based SFQ technology, pulse amplitudes are of order 0.5~mV and pulse widths of order 4~ps. (b) Trajectory on the Bloch sphere for a qubit driven with a resonant SFQ pulse train \cite{McDermott14}. (c) Infidelity of $\left(\pi/2\right)_y$ qubit rotation for resonant (blue) and optimized (red) SFQ pulse sequences versus total sequence length in units of the qubit oscillation period $\tau$. Here we assume 4\% qubit anharmonicity.}
\label{fig:basic}\end{figure}

\indent \indent In spite of prior work to develop SFQ schemes for qubit biasing, until recently there had been no compelling ideas for the coherent control of qubit circuits with SFQ pulses. However, we have recently shown that SFQ pulse trains can be used to induce high-fidelity coherent rotations of the qubit state \cite{McDermott14}. In the simplest implementation, the qubit is irradiated with a train of SFQ pulses with interpulse spacing matched to the qubit oscillation period \cite{Bodenhausen76}. For typical SFQ technology, the pulse duration is of order ps, far shorter than the characteristic qubit oscillation period (e.g., 200~ps for a qubit frequency of 5~GHz). Because the SFQ pulse width is much smaller than the qubit period, the energy deposited per pulse is quite insensitive to the detailed SFQ waveform and is determined rather by the time integral of the pulse, which is precisely quantized to a single flux quantum. As a result, the SFQ pulse can be modeled as a Dirac-$\delta$ function. It is straightforward to show that the energy delivered by a single pulse is given by
\begin{align}
E_1 = \frac{\omega_{01}^2 C_c^2 \Phi_0^2}{2C'};
\label{eq:pulse}
\end{align}
see Fig. \ref{fig:basic}a. Here, $C'$ is the sum of the qubit self-capacitance and the coupling capacitance $C_c$ and the subscript 1 indicates that we refer to the qubit response to a single pulse. For the parameters $\omega_{01}/2\pi$~=~5~GHz, $C$~=~100~fF, and $C_c$~=~100~aF, we find that the single pulse couples an energy to the qubit of order $10^{-4}$ quanta. However, for pulses that are applied coherently (i.e., so that the pulse-to-pulse spacing matches the qubit oscillation period), the energy deposited by the pulse goes as the square of the number of pulses, so that for roughly 100 pulses (corresponding to a sequence length of 20~ns) it is possible to fully excite the qubit.

In more detail, the single SFQ pulse applied to a qubit produces a rotation about a control vector in the equatorial plane of the Bloch sphere with angle
\begin{align}
\delta \theta = C_c \Phi_0 \sqrt{\frac{2 \omega_{10}}{\hbar C}};
\label{eq:angle}
\end{align}
in between pulses, the qubit undergoes free evolution. The SFQ pulse train will induce coherent rotations when the free evolution periods are matched to the oscillation period $2 \pi/\omega_{10}$ of the qubit. For a qubit initially in state $\ket{0}$, the resonant pulse train yields a coherent rotation in the $xz$-plane as depicted in Fig. \ref{fig:basic}b. For a pulse interval that is slightly mismatched from the oscillation period, the state vector slowly drifts away from the $xz$-plane, and in the limit of a large timing mismatch the state vector undergoes small excursions about the north pole of the Bloch sphere.

Potential sources of error in SFQ-based gates are timing jitter of the pulses and weak anharmonicity of the qubit; these have been discussed in detail elsewhere \cite{McDermott14}. Ultimately, the error in SFQ-based control sequences will be dominated by leakage out of the computational subspace. A practical superconducting qubit is not an ideal two-level system \cite{Steffen03}. For a typical transmon qubit \cite{Koch07, Schreier08, Barends13}, the anharmonicity $(\omega_{10}-\omega_{21})/\omega_{10}$ is of order 4-5\%. A single strong SFQ pulse will induce a large spurious population of the $\ket{2}$ state as a result of its broad bandwidth, and leakage errors induced by fast SFQ control pulses have been considered previously \cite{Ohki07}. However, a resonant SFQ pulse train tailored to perform a desired rotation in the 0--1 subspace in a larger number of steps $n$ will show greatly reduced spectral density at $\omega_{21}$, enabling high-fidelity SFQ-based gates with acceptable leakage. We have examined gate fidelity for resonant SFQ pulse trains designed to produce $(\pi/2)_y$ rotations for a range of total numbers of pulses (and hence gate durations); results are shown in Fig. \ref{fig:basic}c (blue trace). Gate errors decrease as $n^{-2}$; by increasing the number of pulses and thus the total duration of the sequence, one reduces the spectral weight of the pulse sequence at the 1--2 transition. For practical transmon qubit parameters, gate fidelity around 99.9\% is achievable for sequence lengths of order 20~ns, compatible with the lengths of conventional microwave-based qubit control sequences.

Moreover, more complex SFQ pulse sequences with variable pulse-to-pulse spacing can provide improvements in gate fidelity for a fixed gate time, with significant benefits in terms of the physical qubit overhead required for robust error correction. Determination of the interpulse intervals that yield the highest fidelity gates is an optimal control problem \cite{Motzoi11}. Criteria for adequate pulse placement include minimization of leakage to higher levels and correct execution of the gate in the computational subspace, as well as robustness against imperfections of the SFQ driver such as pulse timing jitter \cite{Rylyakov99, Bunyk01b}. In this case, standard gradient-based control algorithms are not appropriate for the optimization problem, as during each time step the only options are to apply an SFQ pulse or not, so that differentiation with respect to pulse amplitude is not possible. However, other approaches including those based on genetic algorithms do seem to work. Preliminary work suggests that leakage errors can be suppressed by a factor of 50 for sequence lengths around 20~ns \cite{Liebermann16}. In Fig. \ref{fig:basic}c (red trace) we show the infidelity of an optimized SFQ pulse sequence involving 8 SFQ time steps per qubit oscillation period.

\begin{figure}[t]
\includegraphics[width=.98\columnwidth]{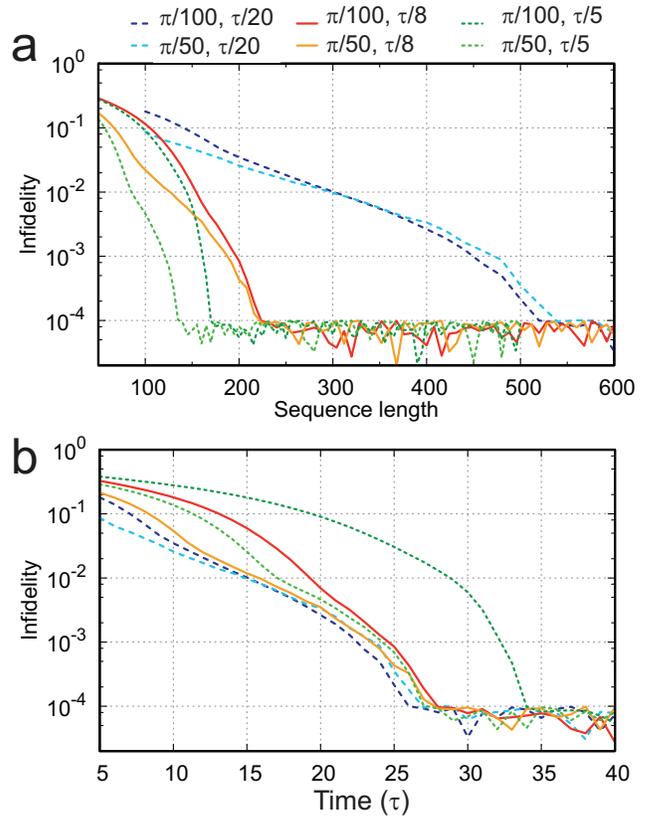}
\vspace*{-0.0in} \caption{Dependence of SFQ gate infidelity on PGU resources. (a) $\left(\pi/2\right)_y$ gate infidelity versus register size for various SFQ tip angles and SFQ timesteps (in units of the qubit oscillation period $\tau$). Here we assume qubit anharmonicity of 4\%. (b) $\left(\pi/2\right)_y$ gate infidelity versus total SFQ sequence length. High frequency clocks require large shift registers. Only for low clock frequency does the required shift register size also depend on the tip angle per SFQ pulse.}
\label{fig:Per}\end{figure}

Ultimately, the design of optimized SFQ-based pulse sequences must be performed with an eye to minimize the resource requirements of the SFQ pulse pattern generator. This leads us to consider the requirements of SFQ register length and clock speed that are needed to attain high gate fidelity. We have performed genetic algorithm-based simulations using the techniques of \cite{Sutton94} with more restrictive assumptions on the SFQ driver. We find that there are tradeoffs between total gate time, which allows drift induced by the undriven part of the Hamiltonian to act; coupling strength of the SFQ driver to the qubit, which sets the timescale for energy transfer; and timing resolution of the sequence, which is set by the SFQ clock frequency. In Fig. \ref{fig:Per}a we plot infidelity of the $\left(\pi/2\right)_y$ gate as a function of register size used to realize the rotation, for various SFQ clock timesteps (in units of the qubit oscillation period $\tau$) and for SFQ tip angles of $\pi/50$ and $\pi/100$; here we assume a qubit anharmonicity of 4\% [e.g., $(\omega_{10}-\omega_{21})/2\pi$~=~200~MHz for $\omega_{10}/2\pi$~=~5~GHz]. For SFQ clock frequency exceeding the qubit frequency by a factor of 8, high-fidelity qubit rotations are achieved with register lengths around 200 bits, with little dependence of gate fidelity on the tip angle provided by the single SFQ pulse. In Fig. \ref{fig:Per}b we plot the same data, converting the horizontal axis to total sequence length in units of $\tau$. Again, for a factor of 8 overhead in SFQ clock frequency compared to qubit frequency, gates approaching the ``quantum speed limit"~$\sim2\pi/(\omega_{10}-\omega_{21})$ set by the qubit anharmonicity are possible.

It is possible to extend these ideas to realize all-SFQ implementations of two-qubit entangling gates. A microwave activated  control-$Z$ (CZ) gate
for two transmon qubits has been previously realized \cite{Chow13}. However, the gate could be implemented with SFQ pulses replacing the microwave drive.
Following \cite{Chow13}, we choose the configuration of transmon energies such that in the absence of interaction, the energy of state $\ket{03}$ coincides with the energy of state $\ket{12}$. When the interaction is turned on, these two levels hybridize and form a split doublet $\ket{\pm} = \left(\ket{12}  \pm \ket{03}\right)/\sqrt{2}$ with energy separation $\sim \sqrt{3} J$, where $J$ is the effective qubit-qubit coupling strength \cite{Majer07}. In this case, the frequencies of transitions $\ket{01}\rightarrow \ket{11}$ and $\ket{11} \rightarrow \ket{+}$ also differ by $J$. We can selectively drive the latter transition by an SFQ pulse sequence containing  $N_2 \gg \omega_{11 \to 12}/J$  pulses. This pulse sequence will cause a full Rabi rotation of state $\ket{11}$  through state $\ket{+}$  back to $\ket{11}$ resulting in an extra phase $\pi$, while other states are not significantly affected by the drive.

%We consider an SFQ-based control-$Z$ ($CZ$) gate applied to a system of two transmon qubits that is based on the cross-resonance scheme considered in \cite{Rigetti10, Chow11}.  We choose the configuration of transmon energies such that in the absence of interaction, the energy of state $\ket{03}$ coincides with the energy of state $\ket{12}$. When the interaction is turned on, these two levels hybridize and form a split doublet $\ket{\pm}$ with energy separation $\sim \sqrt{3}J$, where $J$ is the effective qubit-qubit coupling strength \cite{Majer07}. In this case, the frequencies of transitions $\ket{01}\to \ket{11}$ and $\ket{11}\to \ket{+}$ also differ by $\sim J$. We can selectively drive the latter transition by an SFQ pulse sequence containing $N_2$ pulses, where $2\pi N_2/\omega_{11,21} \gg  1/J$. This pulse sequence will cause a full Rabi rotation of state $\ket{11}$ through state $\ket{+}$ back to $\ket{11}$, while other states are not significantly affected by the drive. It is instructive to compare the gate time we estimate from the above analysis to the gate time associated with passive interaction between qubits.  The latter is characterized by the $CZ$ rate $\zeta_0 \simeq  J^2/\Delta_{11,20}$, and the ratio of gate times for the SFQ-activated gate and the passive gate is $J/\Delta_{11,20}$.  We observe that the gate time is effectively reduced by the factor $\Delta/J \gg 1$. We expect that the extension of optimal control theory to SFQ-based entangling gates will lead to additional reductions in gate time.

For implementations involving the Strauch-type C-phase gate \cite{Strauch03, Kelly15}, the SFQ coprocessor would ideally implement fast baseband flux control in addition to resonant $X$- and $Y$-rotations. As there are detailed descriptions of SFQ-based DACs for the flux control of large-scale arrays in the literature \cite{Johnson10}, we will not attempt a discussion of such efforts here. In order for this approach to be viable for the control of a large-scale surface code, however, the speed and flux resolution of the SFQ-based DACs must be improved significantly over the current state of the art, and the baseband flux controller must be optimized with respect to both dissipation and physical footprint.

\section{V. JPM-based Measurement}

\begin{figure}[t]
\begin{center}
\includegraphics[width=.99\columnwidth]{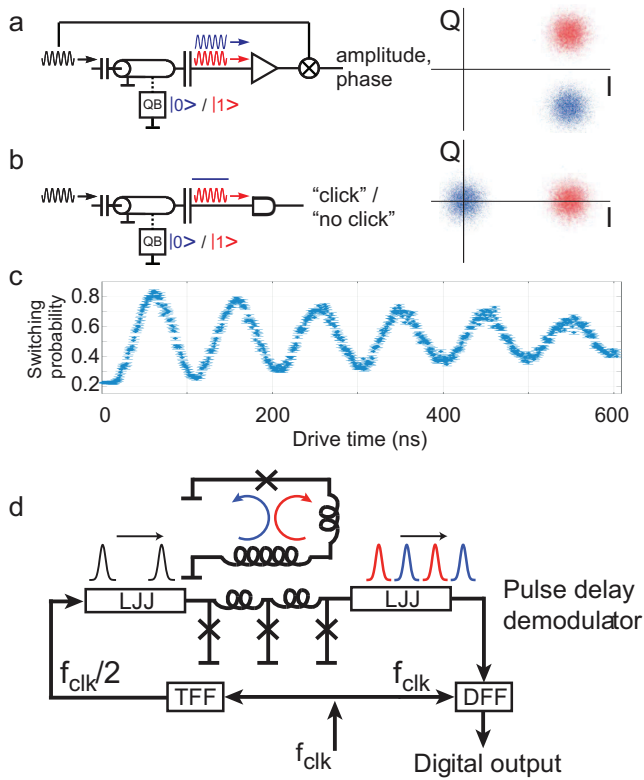}
\vspace*{-0.0in} \caption{(a) Conventional dispersive measurement of a superconducting qubit via heterodyne detection. The state of the qubit is imprinted on the phase of a weak microwave probe tone that is transmitted across a linear resonator dispersively coupled to the qubit. (b) Counter-based qubit measurement with the JPM. Coherent drive at the dressed cavity frequency corresponding to $\ket{1}$ projects the qubit into either $\ket{0}$ or $\ket{1}$ and populates the resonator with a large number of photons $n$ if and only if the qubit is projected into state $\ket{1}$. The JPM interrogates the cavity to determine whether it is in the ``bright" or ``dark" state. (c) JPM-detected qubit Rabi oscillations \cite{Pechenezhskiy16}. (d) JPM readout scheme based on SFQ pulse delay modulation/demodulation. Here the flux state of the JPM rf SQUID loop modulates the delay of an SFQ pulse propagating on an unshunted JTL or long Josephson junction (LJJ). Detection of the delay is done using DFF stage(s) acting as a race arbiter and producing a digital ``1" at even or odd clock periods for non-delayed (blue) or delayed (red) SFQ pulses, respectively. The synchronized digital output is subsequently fed to an SFQ multiplexer for transmission to the classical coprocessor.}
\label{fig:JPM}
\end{center}
\end{figure}

\indent \indent Conventional qubit measurement is based on heterodyne detection of weak microwave probe signals (see Fig. \ref{fig:JPM}a). The approach requires significant cryogenic and room-temperature hardware for analog signal processing and thresholding, and the classical overhead associated with qubit measurement represents a significant obstacle to building towards larger quantum arrays. Our task is to find an efficient means to transfer the classical result of projective quantum measurement to a proximal cryogenic coprocessor for the purpose of error detection and possible postprocessing and feedback. The Josephson Photomultiplier (JPM) \cite{Chen11, Poudel12,Govia12,Govia14b} is an enabling element for the measurement side of the quantum--classical interface, as it provides access to the binary result of projective quantum measurement at the millikelvin stage, without the need for cryogenic amplification or wiring to room temperature in order to perform heterodyning and thresholding. In its simplest implementation, the JPM consists of a Josephson junction biased slightly below the critical current $I_0$. The potential energy landscape $U(\delta)$ for the phase difference $\delta$ across the junction takes on the familiar tilted-washboard form \cite{Tinkham96}, with local potential minima characterized by a barrier height $\Delta U$ and plasma frequency $\omega_p$. The circuit design and bias parameters are chosen so that there are two discrete energy levels in each local minimum of the potential, $\Delta U / \hbar \omega_p \sim 2$; the junction initially occupies the ground state. Microwaves that are tuned to the junction resonance induce a transition to the first excited state, which rapidly tunnels to the continuum. This tunneling transition in turn leads to the appearance of a large voltage across the junction of order twice the superconducting gap. Absorption of a photon thus yields an unambiguous and easily measured ``click".

Several of us have outlined an approach to qubit measurement with the JPM \cite{Govia14}; the basic scheme is shown in Fig. \ref{fig:JPM}b. The qubit (resonating around 5 GHz) is coupled to a readout cavity (resonating around 6 GHz).  As in the usual dispersive limit of the Jaynes-Cummings Hamiltonian, the cavity acquires a dispersive shift $\chi \equiv g^2/\Delta$ that depends on the state of the qubit; here, $g$ is the qubit-cavity coupling rate and $\Delta$ is the qubit-cavity detuning. The measurement proceeds in two stages: (1) First, we map the qubit state to microwave photon occupation of the readout cavity. This can be done by driving the readout resonator at the dressed frequency corresponding to the qubit $\ket{1}$ state for a time equal to $\pi/\chi$. If the qubit is in the $\ket{1}$ state, the microwave drive pulse creates a large photon occupation in the cavity; if the qubit is in the $\ket{0}$ state, however, the cavity acquires a transient occupation but coherently oscillates to a state near vacuum at the end of the drive pulse.  (2) Next, we map photon occupation of the cavity to switching of the JPM (``click'' or ``no click'') by allowing spontaneous emission from the cavity to couple to the JPM. Note that for the ringup portion of the protocol, microwave drive at one of the dressed cavity frequencies can be replaced by irradiation with an appropriate SFQ pulse train; see \cite{McDermott14}.

%This measurement protocol requires microwave drive of the readout resonator at the dressed cavity frequency $\omega_{|1\rangle}$ corresponding to the qubit state $|1\rangle$ for a time $\pi/\chi$.  However, the same result can be achieved by driving the readout cavity with an SFQ pulse train with pulse-to-pulse interval $2\pi/\omega_{|1\rangle}$ and a total number of pulses $N_{\rm pulses} = \omega_{|1\rangle}/2\chi$.  As a result, the average number of photons in the readout resonator is about $n_{\rm ph}\simeq \pi^2 (C_c/C')(\hbar\omega_0 C_c/2e^2)N_{\rm pulses}^2 $ for the qubit in state $|1\rangle$, while $n_{\rm ph}\simeq = 0$ for qubit in state $|0\rangle$; see \cite{McDermott14}.

In Fig. \ref{fig:JPM}c we show data from a typical JPM-detected qubit Rabi scan \cite{Pechenezhskiy16}; in more recent experiments, we have achieved raw single-shot measurement fidelity of 92\% (uncorrected for relaxation and initialization errors) \cite{Opremcak17}. Here, the Josephson junction in the JPM is embedded in an external inductor and the JPM switching event triggers a phase slip in the resulting rf SQUID loop, in analogy to the flux-biased phase qubit \cite{Simmonds04}. The experimental setup involves no isolator or circulator between the JPM and qubit chips; nevertheless, we have shown that by using the intrinsic damping of the JPM to efficiently remove photons generated by the measurement process, we can suppress dephasing associated with measurement backaction even in the absence of bulky nonreciprocal circuit elements.

Crucially, the JPM provides access to the binary classical result of projective quantum measurement at the millikelvin stage of the cryostat: in the case of the flux-biased JPM, the measurement result is stored in the classical circulating current state of the JPM SQUID loop following interaction of the JPM circuit with the qubit readout resonator. As such circulating currents form the basis of Non-Destructive Read Out (NDRO) elements in SFQ digital logic, it is straightforward to convert the result of a JPM-based qubit measurement to a propagating fluxon suitable for subsequent postprocessing by the SFQ-based coprocessor. In one possible implementation, the tunneling transition of the JPM is imprinted on the propagation delay of a fluxon coupled to a proximal Josephson transmission line (JTL) consisting of unshunted, non-dissipative Josephson junctions or a Long Josephson Junction (LJJ).  The use of an unshunted JTL or LJJ ensures dissipationless SFQ propagation, providing for minimal backaction and quasiparticle generation. As discussed in \cite{Semenov03, Fedorov07}, if damping is negligibly small, fluxons can propagate ballistically with a speed depending on the input kinetic energy supplied by the clock generator.  Lateral dc bias current injection along the LJJ can compensate slowing of the fluxon due to viscous drag.

Fig. \ref{fig:JPM}d shows the block diagram of an SFQ circuit designed for readout of the flux state of a JPM. It is based on an SFQ pulse delay modulation/demodulation approach similar to that employed in highly successful low-pass analog-to-digital converters (ADCs) \cite{Rylov97, Mukhanov04}.  In contrast to the typical ADC application, here we need to differentiate only two states of the JPM. As a result, the resolution requirements of the detector are quite modest and a simplified delay demodulator can be used.

In this circuit, the input SFQ clock train is divided into two branches to enable comparison of non-modulated and modulated SFQ pulse streams at the output race arbiter. The toggle flip-flop (TFF) provides a factor of 2 frequency division, and the resulting pulse train is coupled to the probe JTL or LJJ.  The delay-modulated SFQ pulse train is fed to a race arbiter consisting of a single D flip-flop (DFF) or a set of DFFs for more accurate delay differentiation, if necessary. For every modulated SFQ pulse arriving at the DFF data input, the non-modulated clock reads out the DFF state twice, effectively putting the arriving SFQ pulse into either odd or even time bins depending on the induced delay. The resulting phase encoding (Manchester encoding) of the qubit measurement outcome has advantages for the chip-to-chip transmission of the classical result of projective quantum measurement.

In contrast to the C-SQUID readout scheme described in \cite{Semenov03}, our proposed delay-based scheme requires no dissipative SFQ generation/annihilation events during the JPM readout process.  All SFQ circuits based on resistively-shunted junctions (TFF, DFF, splitter) are located on the periphery of the chip remote from the JPM.  Depending on the clock frequency, one can perform repeated measurements of a particular JPM state allowing downstream averaging to achieve higher fidelity readout of the classical circulating current state of the JPM.

We anticipate that the dilute results of JPM-based quantum measurement from up to $\sim$100 channels will be combined on a single line and streamed upward to the SFQ coprocessor, with the address of the syndrome qubit encoded in the timing of the measurement bit (see below). As an alternative to ancilla-based parity measurement in the surface code, however, the JPM is also amenable to direct parity readout, where a single qubit readout resonator is coupled to multiple qubits and the readout resonator drive waveform is tailored to encode parity in cavity photon occupation \cite{Govia15}. Ultimately decisions about whether to perform direct parity readout or ancilla-based parity readout will involve tradeoffs between achievable fidelity and the resource requirements associated with the two approaches.

\section{VI. SFQ-based Coprocessor}

\indent \indent The physical operations needed to realize the surface code can be categorized as follows:

\begin{itemize}

\item {Stabilization of a logical state by repeated syndrome extraction. This mostly involves qubit measurement, reset, Hadamard and CNOT-gates. These operations need to be performed repeatedly and define an error correction cycle.}

\item{Encoding of logical qubits and execution of logical Clifford gates. This mostly involves physical Clifford gates as well as the temporary turn-off of local error correction.}

\item{Magic state distillation and T-gates. This involves the execution of physical T-gates and repeated rounds of error correction.}

\end{itemize}

\begin{figure}[t]
\includegraphics[width=.98\columnwidth]{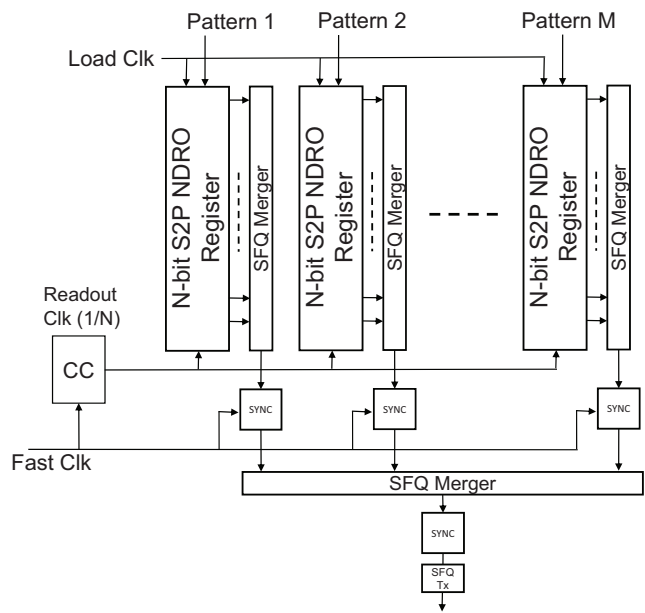}
\vspace*{-0.0in} \caption{Classical SFQ-based PGU. $N$-bit long patterns are loaded from a room-temperature FPGA controller to a set of $M$ $N$-bit NDRO-based serial-to-parallel (S2P) registers. A fast low-jitter clock (e.g., $f_c$ = 40 GHz) is divided by the SFQ clock controller (CC) to produce an $f_c/N$ readout clock used to read the loaded pattern from the $N$-bit NDRO registers.  The pattern goes to SYNC for synchronization with the low-jitter global clock. The SFQ merger directs the pattern to the output SFQ transmitter. }
\label{fig:PGU}\end{figure}

An interesting feature of the surface code is that on the physical level only a relatively small set of gates is required, corresponding to a limited set of SFQ control registers. The realization of complex bit streams for high-fidelity qubit control demands a special-purpose SFQ processor, a Pattern Generator Unit (PGU) capable of producing trains of SFQ voltage pulses with variable interpulse delay.  Figure \ref{fig:PGU} shows one possible implementation. The PGU is based on an array of SFQ shift registers with serial input and parallel output (S2P) constructed using NDRO elements.  The qubit SFQ control patterns are serially pre-loaded to the individual S2P registers using an SFQ load clock from memory (either room-temperature or cryogenic RAM).  As a result, each S2P register will hold either a partial or a complete SFQ pulse pattern with the required interpulse spacing set by the number of 0s in the register cell.

The readout SFQ clock (Fast Clk) performs parallel non-destructive readout from the S2P registers in a wave-pipeline fashion starting from the last bit.  This action creates an SFQ pulse train that replicates the pre-loaded bit pattern. A synchronizer based on D flip-flops ensures accurate timing of the bit patterns read out from individual S2P registers.

For longer control sequences that must be stored across multiple S2P registers, it is necessary to merge the various pieces into a complete bit pattern for serial streaming to the qubit. The SFQ merger combines bit patterns from individual S2P registers into a single bit stream. Since the merger is an asynchronous device, stitching the individual patterns from different S2P elements will be challenging.  For this purpose, SFQ timing gates (SYNC) are added to the readout clock port of each S2P register in order to synchronize the readout of the various registers.  The combined SFQ bit stream is re-synchronized at the merger output prior to transmission to the qubit chip. Alternatively, the SFQ merger could be replaced by a shift register with parallel input and serial output (P2S).  This arrangement would ensure full synchronicity of the readout pulse train. Compared to the implementation involving an SFQ merger and output synchronizer gates, the P2S readout register would involve $3N-4$ additional junctions, where $N$ is the number of bits in the segment.

The NDRO-based registers would allow local storage of control bit patterns for repeated streaming, as needed, e.g., for randomized benchmarking, process tomography, or repeated implementation of the surface code cycle. In a way, the PGU acts as an operational memory, storing the most frequent bit patterns and bit patterns required for the next quantum operation steps.  During operation of an algorithm, updating of the PGU register contents is required only for a fraction of the S2P registers. Significant reductions in the power dissipation and footprint of the PGU could be obtained by recycling baseline bit sequences and using relatively dilute patterns of bits to ``fine tune" control sequences for individual qubits, whose transition frequencies and anharmonicities will in practice differ due to inevitable disorder in the qubit Josephson energies \cite{Rosenblatt17, Hutchings17}. This could be accomplished using smaller shift registers tailored to the high-fidelity control of individual qubits. The dilute arrays of correction bits would be merged with dense baseline bit patterns using SFQ-based XOR gates for the purpose of suppressing leakage errors. Alternatively, it might be possible to reduce the required register size by implementing control sequences that consist of shorter bit patterns that are streamed repeatedly to the qubit with appropriate interword delays tailored to minimize gate error.

\section{VII. Vision and Challenges}

\indent \indent We envision a scheme where the SFQ coprocessor is operated at the cold stage ($T\sim3$~K) of a pulse tube cooler and coupled to the quantum array via low-loss microstrip lines. For the coprocessor we anticipate using conventional high-$J_c$ Nb-AlO$_x$-Nb junctions with critical current density of order 1~kA/cm$^2$, corresponding to junction critical currents of order 100~$\mu$A. A single SFQ junction with critical current 100~$\mu$A undergoing phase slips at an average rate of 5~GHz will dissipate power of order 1~nW. For each qubit, we store a control waveform consisting of $10^3$ bits, corresponding to a control sequence length of 30~ns, assuming a clock frequency of 30~GHz. The power dissipation per channel in the PGU is then 0.1~$\mu$W if we assume a 10\% duty cycle per channel, resulting in a total power dissipation of 10~W for a PGU capable of delivering unique, independent bit patterns to a quantum array consisting of $10^8$ qubits. This power dissipation is well in line with the cooling power available for state of the art pulse tube coolers, where single units achieve cooling power up to 2~W at temperatures of 4~K. For a special-purpose cryostat designed to support a large-scale multiqubit array, it will be straightforward to operate several such pulse tube units in parallel.

We anticipate the need for a quantum--classical interface chip to mediate the interaction between the quantum array and the classical coprocessor. The dissipative interface chip is coupled in a flip-chip arrangement to the quantum chip to form a multichip module (MCM). In this scheme, the SFQ pulses streamed from the PGU are communicated to the quantum array via  capacitive coupling across the chip-to-chip gap. For example, a coupling electrode with area of order $10\times10$~$\mu$m$^2$ will be positioned directly over the transmon island and the vacuum gap between the two chips of order 10~$\mu$m will provide a coupling capacitance of order 100~aF, ideal for the realization of high-fidelity SFQ-based control sequences. In this scenario, no SFQ junctions are required on the quantum chip, so that the classical and quantum fabrication processes are completely decoupled; the modular approach to fabrication yields a significant simplification compared to efforts at monolithic integration of SFQ and qubit elements on a single chip. The interface chip will involve qubit readout resonators, JPM detectors, SFQ-based transmit/receive elements for communicating bit patterns to the qubit and converting JPM measurement results into propagating fluxons, and multiplexers/demultiplexers (MUX/DEMUX) for measurement and control, respectively. The MUX/DEMUX elements will be required to streamline the wiring between the quantum array and the coprocessor over low-thermal conductance, high-bandwidth microstrip lines, as described below. In previous work, the signal distribution challenge has been explored in the context of SFQ-controllable microwave switches and filters \cite{Rafique09, Naaman16}. For the schemes we propose here, the control bit patterns are dense and data-rich compared to the dilute results of syndrome measurement. As a result, the interface chip will house a larger number of DEMUX elements than MUX elements to support a given number of qubits. As an estimate, here we assume a single 100-bit MUX and ten 10-bit DEMUX channels serving 100 qubits.

\begin{figure}[t]
\includegraphics[width=.98\columnwidth]{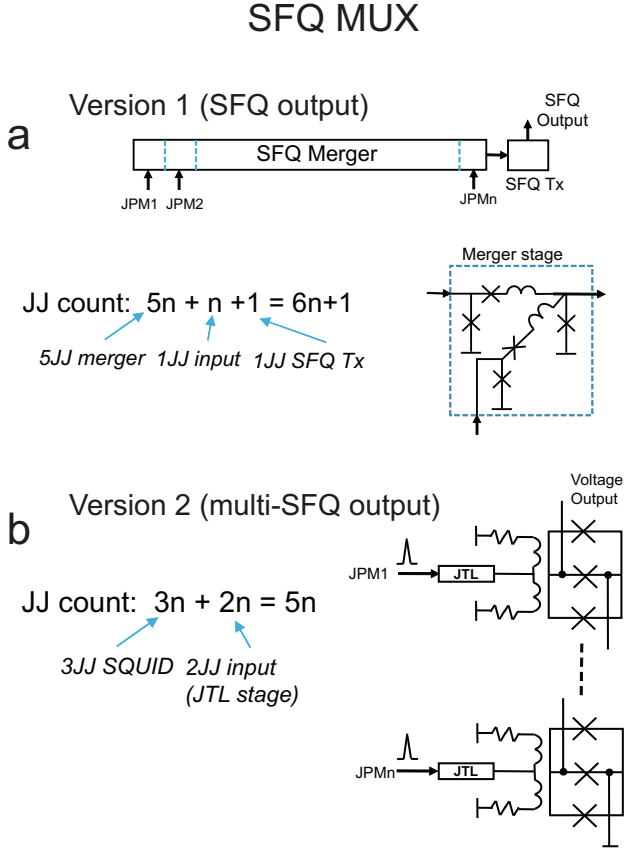}
\vspace*{-0.0in} \caption{SFQ MUX to merge JPM measurement results for transmission from the interface chip to the SFQ PGU. (a) Version 1, based on SFQ mergers. (b) Version 2, based on SQUID stacks.}
\label{fig:MUX}\end{figure}

There are two possible versions of the readout MUX. One is based on a simple SFQ-merger JTL bus channeling data from the individual JPMs to a single output node for subsequent SFQ pulse transmission to the 3~K coprocessor (see Fig. \ref{fig:MUX}a).  This arrangement assumes sequential (non-simultaneous) qubit readout for the channels addressed by the MUX. For a 100-bit MUX, this would require $\sim 5\times100 + 100 =600$~junctions total, corresponding to 6 junctions per readout channel (here we assume 5 SFQ junctions per SFQ merger in the MUX, with an additional overhead of 100 SFQ junctions). Another option is to use a SQUID stack to produce a multi-SFQ pulse for transmission (see Fig. \ref{fig:MUX}b).  The state of the SQUID will be controlled by JPM readout channels that are inductively coupled to each SQUID of the stack. This arrangement would require only $100 \times 2 + 100 = 300$~junctions, but it is slower compared to the SFQ merger-based MUX.

\begin{figure}[t]
\includegraphics[width=.98\columnwidth]{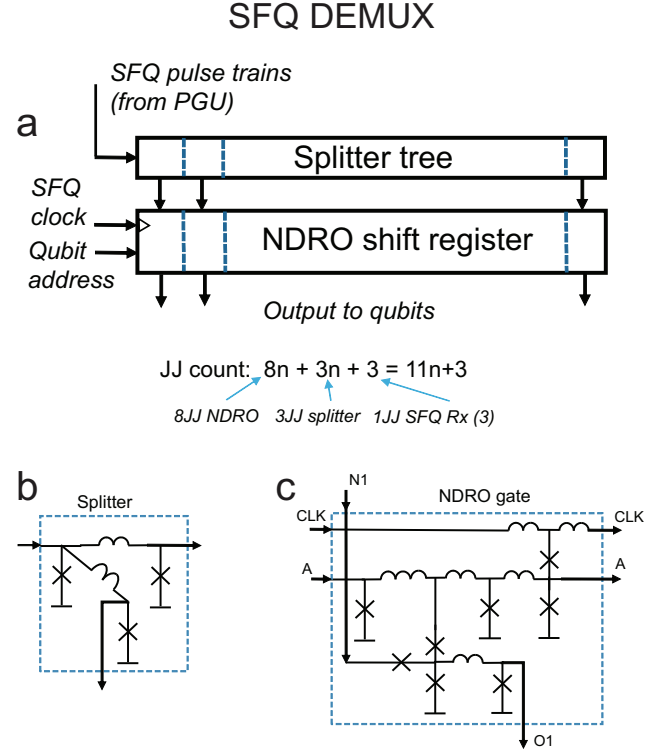}
\vspace*{-0.0in} \caption{SFQ DEMUX to distribute qubit control at the interface chip. (a) Block diagram of the DEMUX, along with circuit diagrams of (b) the SFQ pulse splitters and (c) NDRO gates used in (a).}
\label{fig:DEMUX}\end{figure}

The control DEMUX can be implemented as an SFQ splitter tree, in which each branch is controlled by an NDRO gate containing the address (see Fig. \ref{fig:DEMUX}). For a 10-bit DEMUX, this would require $\sim 3 \times 10 + 8 \times 10 + 3 = 113$~junctions, or approximately 11 junctions per control channel (here we assume 3 SFQ junctions per SFQ splitter; 8 SFQ junctions per NDRO gate; and an overhead of 3 SFQ receiver junctions per DEMUX cell). In the case of sequential control functions, the address is simply a token bit shifted between NDRO cells.  If SFQ control functions need to be applied out of order, a DEMUX programming address line would be required.  Alternatively, a qubit address header could be prefixed to each control SFQ pulse train, requiring a more complicated addressable switch in place of the DEMUX.

For the sake of an estimate of the power budget at the interface chip, we thus assume of order 20 SFQ JJs per qubit channel. At the millikelvin operation temperature it is possible to realize these junctions using a low-$J_c$ technology corresponding to junction critical currents of order 1~$\mu$A. We again assume a qubit operation frequency of 5~GHz and a 10\% operational duty cycle. We find an average power dissipation at the interface chip of 20~pW per qubit channel, corresponding to a total power dissipation of 2~mW for a quantum--classical interface matched to a $10^8$-element quantum array. This level of dissipation is compatible with the cooling power available at the millikelvin stage from a state of the art dilution refrigerator.

The design of a scalable system must be informed by considerations of physical footprint as well as power dissipation. The superconducting qubit cell occupies roughly $100\times100$~$\mu$m$^2$. It is necessary that the control and measurement hardware on the interface chip match this physical footprint. We envision an architecture where each qubit in the quantum array is coupled capacitively to a compact readout resonator on the interface chip; a proximal photon counter is used to probe the photon occupation of the readout mode. The physical footprints of the inductively-biased JPM and of a compact, lumped-element $LC$ readout resonator are well-matched to the $\sim$100~$\mu$m lateral cell dimension of the quantum array. Alternatively, we expect that it will be possible to reduce the number of required JPM channels by employing a hybrid time/frequency-domain multiplexing scheme for readout, wherein a single JPM is coupled sequentially via microwave sideband pulses \cite{Strand13} to a handful of readout resonators (say, four) operating at slightly different frequencies. As discussed above, the reduction in power obtained by moving to lower $J_c\sim10$~A/cm$^2$ involves an increase in the physical footprint of the SFQ elements due to the larger inductances $\sim100$~pH associated with the low-power SFQ technology. However, it is possible to realize 2- or 3-turn inductors of the right magnitude with a lateral footprint of order 10~$\mu$m; if needed, these inductors could be fabricated as compact nanowires from high-kinetic inductivity disordered superconducting films or as compact Josephson junction arrays to ensure the inductor self resonance falls far outside the SFQ pulse bandwidth. We expect the $\sim 20$ SFQ junctions required for readout and control of each qubit to occupy an area of order $20\times100$~$\mu$m$^2$, quite a bit smaller than the footprint of the qubit itself. The dissipation at the interface chip corresponds to a power density of 200~nW/cm$^2$, so efficient cooling of the interface chip and thermal decoupling of the interface chip from the quantum array will be critical. Ultimately, the design of the interface chip will involve tradeoffs between physical footprint and dissipation.

We now consider the thermal budget and footprint of the wiring. We envision a scenario where classical SFQ bitstreams are transmitted between the interface chip and the PGU on low-loss superconducting microstrip transmission lines fabricated on flexible Kapton tape. There have been prior demonstrations of the high-fidelity transmission of SFQ pulses over centimeter scales \cite{Polonsky93}, and the fabrication of superconducting microstrips on flex lines is an established technology \cite{vanWeers12}. For a given Kapton thickness, considerations of heat load and wiring footprint favor higher stripline impedance, so we consider the case of 50~$\Omega$ striplines, which are a good match to low-$J_c$ junctions that are favored at low temperature. For 0.5~mil (13~$\mu$m) Kapton, a 50~$\Omega$ impedance is achieved for a trace width around 50~$\mu$m. We assume a conservative trace-to-trace spacing of 50~$\mu$m. We separately consider the heat load of the Kapton dielectric and of the superconducting traces. As discussed above,we assume a modest amount of multiplexing at the level of the interface chip, so that communication with the $10^8$ physical qubit array is accomplished with around $10^7$ control lines.

For Kapton HN, the thermal conductivity in the millikelvin range has been measured to be \cite{Lawrence00}
\begin{align}
\kappa_{\rm Kap} = 4.6\times10^{-3} \, \left(\frac{T}{\rm K}\right)^{0.6} \rm W/m \, K.
\end{align}
The total cross-sectional area of the Kapton wiring is $1.3 \times 10^{-2}$~m$^2$, and we assume a 1~m length from the 3~K pulse tube stage to the millikelvin stage; to get a worst-case idea of the thermal budget, we assume that no effort is taken to heat sink the wires at the still or intermediate cold plate of the dilution refrigerator. We find a total heat load due to the Kapton alone of 220 $\mu$W.

Next, we consider the heat load due to the metal traces of the microstrip ground plane and signal lines. We will use a superconducting material as far below $T_c$ as possible, so that the electronic contribution to thermal conductivity is suppressed. Superconducting Nb would be a natural choice; however, the phonon contribution to thermal conductivity in superconducting Nb is rather large \cite{Kes74}, so Nb is not suitable. NbTi alloy, however, has excellent thermal properties, and NbTi lines have previously been sputtered on flexible Kapton substrates for wiring to superconducting devices \cite{Cantor17}. The low-temperature thermal conductivity of NbTi has been characterized as \cite{Olson93}
\begin{align}
\kappa_{\rm NbTi} = 0.027 \, \left(\frac{T}{\rm K}\right)^2 \rm W/m \, K.
\end{align}
If we assume 100~nm-thick NbTi traces for the signal lines and groundplanes of the microstrips, we find a cross-sectional area of $1.5\times10^{-4}$~m$^2$ for the $10^7$ control lines, leading to a heat load from 3~K to the millikelvin stage of 40~$\mu$W (again, assuming no effort to heat sink the microstrip lines at intermediate temperature stages). We see that the total thermal budget of the 10$^7$ control lines is of order 300~$\mu$W, well within the capacity of a large-scale dilution unit.

By comparison, we can consider the heat load on the cryostat for a conventional heterodyne-based readout and microwave control implementation with $10^8$ qubits. Although the wiring heat load from 3~K to the millikelvin stage will be comparable to that associated with an SFQ-based implementation, the dissipation associated with amplifier bias is especially problematic in the case of conventional heterodyne measurement. We assume a generous multiplexing capability of 100 qubits per superconducting amplifier, and we make the assumption (also generous) that 100 superconducting amplifiers can be read out with a single HEMT-based postamplifier at the 3~K stage. As the drain-source bias of the HEMT dissipates approximately 10~mW, the $10^4$ amplifiers needed to monitor the quantum array will dissipate 100~W, an order of magnitude larger than the estimated dissipation of the SFQ-based PGU described above (of course this estimate does not take into account dissipation in the sophisticated switching matrix that would be needed to multiplex $10^4$ qubit measurement signals onto a single HEMT channel). We assume that some variant of the TWPA is used as the first-stage amplifier. Here, a strong parametric pump tone is required to bias the amplifier in the active state. For state of the art TWPAs, delivery of the pump tone to the amplifier dissipates approximately 100~nW at the millikelvin stage \cite{Hover16}. For $10^6$ TWPAs, we thus find a total power dissipation of 100~mW at the millikelvin stage, far beyond the cooling capacity of the most powerful dilution refrigerator that we can envision by scaling up from present-day technology.

\section{VIII. Conclusion and Future Directions}

To conclude, we have proposed a vision for integration of a large-scale superconducting quantum array with a classical coprocessor based on the Single Flux Quantum digital logic family. The coprocessor is well-matched in terms of physical footprint to a quantum array consisting of up to $10^8$ transmon qubits,  and the associated power dissipation and wiring heatload are compatible with the cooling power available from a large-capacity dilution refrigerator. The approach promises major reductions in system footprint, latency, and dissipation compared to the current state of the art for pulsed microwave coherent control and heterodyne detection of the qubit state. While many of the required technologies are highly developed separately, the integration of the necessary pieces into an optimized system is technically demanding and significant challenges remain. Specifically:

\begin{itemize}

\item{It is essential to demonstrate robust operation of complex, large-scale SFQ processors with high integration density. While most of the elements needed for the development of the SFQ-based PGU have been demonstrated previously, a coprocessor on the scale of what we describe here would be quite novel, though not a qualitative leap from prior works.}

\item{The proposed scheme relies on the robust, low-jitter transmission of SFQ pulses across microstrip flex interconnects connecting the various stages of a vacuum cryostat. While SFQ transmission over microstrip lines and the necessary interconnect technology are both well established, these two pieces have never been combined together. It is necessary to develop appropriate technology for SFQ pulse timing synchronization across a large-scale system consisting of multiple modules and to optimize the design and number of repeater stages for robust transmission of classical data.}

\item{It is necessary to minimize the dissipation and maximize fidelity of JPM-based readout and to demonstrate robust SFQ-based coherent control of qubits. Initial demonstrations reveal raw JPM-based measurement fidelity of 92\% (uncorrected for initialization and relaxation errors) \cite{Opremcak17} and SFQ-based Rabi oscillations have now been shown \cite{Leonard17}. It is necessary to increase JPM-based measurement fidelity and to demonstrate SFQ-based coherent control well beyond the fault-tolerant threshold.}

\item{Optimization of the interface chip will require a match between the footprint of the readout and control elements and the footprint of the qubits. It will be necessary to develop more compact JPM elements and readout resonators. It is essential to ensure that flip-chip integration with the quantum array does not degrade qubit coherence. Our proposed scheme relies on the dissipation of a modest amount of power near the quantum array. While quasiparticle traps are highly effective \cite{Patel16}, it is critical to ensure that operation of the SFQ elements on the interface chip does not degrade qubit performance.}

\item{Optimization of MUX/DEMUX elements at the interface stage is needed to streamline wiring and provide maximum functionality in an efficient, low-power manner. It is necessary to minimize both the footprint and power dissipation of SFQ elements at the interface stage; here, the integration of compact kinetic inductors with low-$J_c$ SFQ junctions would be highly useful.}

\item{System identification and calibration of a large scale quantum array will require new approaches \cite{Stenberg14}. It is necessary to devise efficient approaches to the automated tuneup of optimal SFQ-based control sequences for qubit arrays subject to disorder in the qubit energies and interaction strengths. Gate design must be performed with an eye to the available resources of the SFQ-based PGU (clock frequency, register length, etc.).  An interesting question in this context is how many classical bits are needed for the coherent control of a qubit to a given level of fidelity.}

\item{For implementations based on C-phase Strauch gates \cite{Strauch03}, it will be necessary to develop improved approaches to high-fidelity baseband control using SFQ logic. Prior demonstrations of SFQ-based flux control for large arrays have involved relatively slow, low-resolution DACs and dissipation in the classical controller limited experimental repetition rates \cite{Johnson10}.  Alternative approaches that rely exclusively on cross-resonance gates would need local flux control only to fine-tune the qubit operating point; in this case there is negligible dissipation associated with setting the qubit bias and the demands on the flux controller are rather modest.}

\item{Although the scheme we have outlined here results in significantly smaller dissipation at the millikelvin and 3~K stages of the cryostat compared to conventional microwave control and heterodyne readout, these heat loads are still somewhat beyond the available cooling power of the most powerful dilution refrigerators that are currently available commercially. Nonetheless, there is a straightforward path to achieving cooling powers of a few mW at the millikelvin stage by operating multiple ($\sim$3-10) dilution units in parallel. In addition, 10s of W of cooling power at the 3~K stage could be achieved by running multiple ($\sim$10) pulse-tube refrigerators in parallel on the 3~K stage. This configuration would not be practical or affordable for a cryostat in a typical research lab, but would be entirely feasible for a dedicated large-scale quantum processor.}

\end{itemize}

Despite these challenges, we believe that there are clear directions for continued progress, and the benefits of tight integration of a high-speed, low-power SFQ-based classical coprocessor with a large-scale quantum array justifies intensive work in this area.

\section{Acknowledgments}

R.M., B.L.T.P., M.G.V., and F.K.W. acknowledge support by the U.S. Government under Grant W911NF-15-1-0248, and R.M. and B.L.T.P. acknowledge funding from the NSF under Grants QIS-1720304 and QIS-1720312, respectively.

\end{document}